\documentclass{article}

\usepackage{graphicx}
\usepackage{amsmath}

\title{Essay:\\Construction Theory, Self-Replication, and the Halting Problem}

\author{Hiroki Sayama\\Department of Bioengineering\\
Binghamton University, State University of New York\\
P.O. Box 6000, Binghamton, NY 13902-6000\\Tel: 607-777-4439\\
Fax: 607-777-5780\\Email: sayama@binghamton.edu}

\date{}

\begin{document}
\maketitle

\begin{abstract}
This essay aims to propose {\em construction theory}, a new domain of
theoretical research on machine construction, and use it to shed light
on a fundamental relationship between living and computational
systems. Specifically, we argue that self-replication of von Neumann's
universal constructors holds a close similarity to circular
computational processes of universal computers that appear in Turing's
original proof of the undecidability of the halting problem. The
result indicates the possibility of reinterpreting a self-replicating
biological organism as embodying an attempt to solve the halting
problem for a {\em diagonal} input in the context of
construction. This attempt will never be completed because of the
indefinite cascade of self-computation/construction, which accounts
for the undecidability of the halting problem and also agrees well
with the fact that life has maintained its reproductive activity for
an indefinitely long period of time.
\end{abstract}


\newpage

\section{Introduction: von Neumann's automata and construction theory}

\begin{figure}[t]
\begin{center}
\includegraphics[width=0.4\columnwidth]{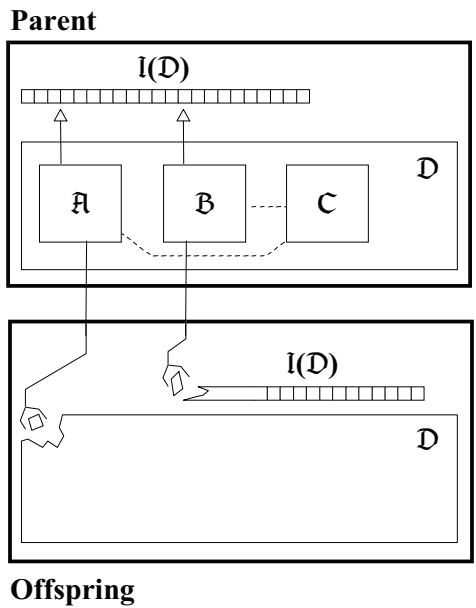}
\end{center}
\caption{Schematic illustration of von Neumann's self-reproducing
automaton. It consists of three parts (universal constructor $\cal A$,
tape copier $\cal B$, and controller $\cal C$; $\cal D \equiv A+B+C$)
and an instruction tape $\cal I(D)$. Since $\cal I(D)$ contains the
information about how to construct the automaton $\cal D$ itself, the
whole system $\cal D+I(D)$ can self-replicate.}
\label{neumann}
\end{figure}

John von Neumann's theory of self-reproducing automata
\cite{vonneumann1951,vonneumann1966} is now regarded as one of the
greatest theoretical achievements made in early stages of artificial
life research \cite{marchal1998,sipper1998,mcmullin2000}. Before
working on its specific implementation on cellular automata, von
Neumann sketched a general outline of his self-reproducing automaton
that consists of the following parts \cite{vonneumann1951}:
\begin{description}
\item{$\cal A$:} A universal constructor that constructs a product
$\cal X$ from an instruction tape $\cal I(X)$ that describes how to
construct $\cal X$.
\item{$\cal B$:} A tape copier that duplicates $\cal I(X)$.
\item{$\cal C$:} A controller that dominates $\cal A$ and $\cal B$ and
does the following:
\begin{enumerate}
\item Give $\cal I(X)$ to $\cal A$ and let it construct $\cal X$.
\item Pass $\cal I(X)$ to $\cal B$ and let it duplicate $\cal I(X)$.
\item Attach one copy of $\cal I(X)$ to $\cal X$ and separate $\cal
X+I(X)$ from the rest.
\end{enumerate}
\end{description}
The functions of these parts are symbolically written as
\begin{eqnarray}
&& \cal A + I(X) \to A + I(X) + X , \\
&& \cal B + I(X) \to B + {\rm 2} I(X) , \\
&& \cal (A + B + C) + I(X) \nonumber \\
&& \cal ~~~~~~~ \to ((A + I(X)) + B + C) \nonumber \\
&& \cal ~~~~~~~ \to ((A + I(X) + X) + B + C) \nonumber \\
&& \cal ~~~~~~~ \to (A  + (B + I(X)) + C) + X \nonumber \\
&& \cal ~~~~~~~ \to (A  + (B + {\rm 2} I(X)) + C) + X \nonumber \\
&& \cal ~~~~~~~ \to (A + B + C) + I(X) + X + I(X) \nonumber \\
&& \cal ~~~~~~~ \to \left\{ (A + B + C) + I(X) \right\}
 + \left\{ X + I(X) \right\} . \label{roleofC}
\end{eqnarray}
Then self-replication can be achieved if one lets $\cal X=D \equiv A+B+C$,
i.e.,
\begin{eqnarray}
\cal D + I(D) &\to&
\cal \left\{ D + I(D) \right\} + \left\{ D + I(D) \right\} . \label{selfrep}
\end{eqnarray}
Figure \ref{neumann} illustrates these notations visually. Note that
for the above conclusion to apply, $\cal D$ must be within the product set
of $\cal A$, which is by no means trivial.

Alan Turing's preceding work on computationally universal machines
\cite{turing1936} gave a hint for von Neumann to develop these
formulations of self-reproducing automata, especially on the idea of
universal constructor $\cal A$. These two kinds of machines apparently
share a similar concept that a universal machine, given an appropriate
finite description, can execute arbitrary tasks specified in the
description. We should note however, that this similarity has often
been overstated in the literature, leading to some misunderstandings
of von Neumann's original intention, recently argued by McMullin
\cite{mcmullin2000,mcmullin1993}. The most significant difference
between these two types of universal machines is that the
constructional machine must be made of the same parts that it operates
on, and therefore both the machine and the parts must be embedded in
the same space-time and obey the same ``physical'' rules, while the
computational machine can be separate from the symbols it operates on,
like the Turing machine's head that exists outside its tape. Another
equally important difference is that computational universality is
defined by the ability of {\em computing the behavior of all the other
models of computation}, while the constructional universality is
defined by the ability of {\em constructing all the structures in a
given specific product set}, which has nothing to do with the ability
of computing the behavior of other constructors. The latter issue will
be revisited later.

The aforementioned differences suggest the need for a distinct domain
of research specially dedicated to the issues of machine construction,
pioneered by von Neumann's work on constructional machines but since
left unnamed to date. This would be closely related to computation
theory pioneered by Turing's work, but should be unique by involving
physical interpretation and implementation of production processes and
thereby connecting logic and mathematics to biology and
engineering. Here I propose to call it {\em construction theory}, a
domain of research that focuses on the theoretical aspects of
productive behaviors of natural or artificial systems, including
self-replication, self-repair and other epigenetic processes. There is
a recent resurgence of studies on these topics in artificial life and
other related fields
\cite{freitas2004,zykov2005,buckley2006,ewaschuk2006,suzuki2006,zhang2006}. Like
in computation theory, there are many important problems yet to be
addressed in construction theory, such as identifying the class of
constructible structures with a given set of physical rules; obtaining
necessary/sufficient conditions for a universal constructor to exist
for a given product set; determining whether there is a single {\em
truly universal} construction model that could emulate all other
construction models; etc.

In what follows, we will focus on one particular question regarding
the relationship between computation and construction theories. While
von Neumann's universal constructor was largely inspired by Turing's
universal computer, what the entire self-replicating automaton $\cal
D$ in construction theory would parallel in computation theory
remained unclear to many, perhaps because von Neumann himself did not
detail in his writings how his theoretical model was related to
computation theory. Besides the universal constructor $\cal A$, the
automaton $\cal D$ also includes $\cal B$ that duplicates a given tape
and $\cal C$ that attaches a copy of the duplicated tapes to the
product of $\cal A$. They are the subsystems that von Neumann added to
the automaton in view of self-replication (and subsequent evolutionary
processes). Their counterparts are not present in the design of Turing
machines, and therefore, the entire architecture of self-reproducing
automata has often been considered a heuristic design meaningful only
on the construction side, but not on the computation side.

Here I would like to help readers realize that self-replication in
construction theory actually has a fundamental relationship with the
diagonalization proof of the undecidability of the halting problem in
computation theory. This relationship was already suggested by some
mathematicians and theoretical computer scientists
\cite{myhill1964,rogers1967}; however, it somehow failed to bring a
broader conceptual impact to other related fields, including
theoretical biology, artificial life, and complex systems
research. Specifically, the mathematical description of
self-replication by von Neumann's universal constructors is of
identical form with the circular computational processes of universal
computers that appear in Turing's original proof of the undecidability
of the halting problem. This leads us to a new interpretation of a
self-replicating biological organism as embodying an attempt to solve
the undecidable halting problem for a {\em diagonal} input, not in
computation theory but in the context of von Neumann's construction
theory. This attempt, of course, will never be completed in a finite
time because of the indefinite cascade of
self-computation/construction, which accounts for the undecidability
of the halting problem and also agrees well with the fact that life
has maintained its reproductive activity for an indefinitely long
period of time.

\section{The halting problem}

The halting problem is a well-known decision problem in theoretical
computer science that can be informally described as follows:
\begin{quote}
{\em Given a description of a computer program and an initial input it
receives, determine whether the program eventually finishes
computation and halts on that input.}
\end{quote}
This problem has been one of the most profound issues in computation
theory since 1936 when Turing proved that no general algorithm
exists to solve this problem for any arbitrary programs and inputs
\cite{turing1936}. The conclusion is often paraphrased that the
halting problem is {\em undecidable}.

Turing's proof uses {\em reductio ad absurdum}. A well-known
simplified version takes the following three steps.

First, assume that there is a general algorithm that can solve the
halting problem for any program $p$ and input $i$. This means that
there must be a Turing machine $M$ that always halts for any $p$ and
$i$ and computes the function
\begin{equation}
f(p,i) \equiv
	\begin{cases}
	1 & \text{if the program $p$ halts on the input $i$,} \\
	0 & \text{otherwise.}
	\end{cases}
\end{equation}

Second, one can easily derive from this machine another Turing machine
$M'$ whose behavior is modified to compute only {\em diagonal}
components in the $p$-$i$ space, i.e.,
\begin{equation}
f'(p) \equiv f(p,p) . \label{diagonal}
\end{equation}
This machine determines whether the program $p$ halts when its
self-description is given to it as an input. Such self-reference would
be meaningless for most actual computer programs, but still
theoretically possible.

Then, finally, one can tweak $M'$ slightly to make yet another machine
$M^*$ that falls into an infinite loop if $f'(p) = 1$. What could
happen if $M^*$ was supplied with its self-description $p(M^*)$?  It
eventually halts if $f'(p(M^*)) = f(p(M^*),p(M^*)) = 0$, i.e., if it
does not halt on $p(M^*)$. Or, it loops forever if $f'(p(M^*)) =
f(p(M^*),p(M^*)) = 1$, i.e., if it eventually halts on $p(M^*)$. Both
lead to contradiction. Therefore, the assumption we initially made
must be wrong---there must be no general algorithm to solve the
halting problem.

\section{Turing's original proof}

Here I would like to bring up an informative yet relatively untold
fact that Turing himself did not like to have such a tricky
mathematical treatment as the above third step that introduces a
factitious logical inversion into the mechanism of the machine, so he
intentionally avoided using it in his original proof. Below is a quote
from his original paper \cite[p.246]{turing1936}, which tells us how
unique Turing's thought was and how much emphasis he placed on an
intuitive understanding of mathematical concepts:
\begin{quote}
{\em ``... The simplest and most direct proof of [the fact that there
is no general process for determining whether a given program
continues to write symbols indefinitely] is by showing that, if this
general process exists, then there is a machine which computes
$J$\footnote{A binary sequence whose $n$-th digit is a Boolean {\em
inverse} of the $n$-th digit of the $n$-th computable sequence. If
this sequence is computable, then it must be listed somewhere in the
series of the computable sequences, which however causes a
contradiction because its diagonal element must be both 0 and 1 at the
same time. Therefore this sequence cannot be computable.}. This proof,
although perfectly sound, has the disadvantage that it may leave the
reader with a feeling that ``there must be something wrong''.  The
proof which I shall give has not this disadvantage, ...''}\\ (Footnote
added by the author)
\end{quote}
Instead, what he actually did for the proof was summarized in the
following paragraph \cite[p.247]{turing1936}:
\begin{quote}
{\em ``... Now let $K$ be the D.N\footnote{Description Number: An
integer that describes the specifics of a given computational
machine.} of $H$\footnote{A machine that incrementally and indefinitely
computes the diagonal sequence of the infinite matrix made of all the
infinitely long computable sequences enumerated in the order of D.N's
of corresponding machines.}. What does $H$ do in the $K$-th section of
its motion? It must test whether $K$ is satisfactory\footnote{An
integer $N$ is considered satisfactory if the machine whose D.N is $N$
can keep writing symbols indefinitely without falling into a deadlock
(Turing called this property {\em circle-free}).}, giving a verdict
``$s$'' or ``$u$''. Since $K$ is the D.N of $H$ and since $H$ is
circle-free, the verdict cannot be ``$u$''. On the other hand the
verdict cannot be ``$s$''. For if it were, then in the $K$-th section
of its motion $H$ would be bound to compute the first $R(K-1)+1 =
R(K)$\footnote{$R(N)$ denotes how many machines are circle-free within
the range of D.N's up to $N$.} figures of the sequence computed by the
machine with $K$ as its D.N and to write down the $R(K)$-th as a
figure of the sequence computed by $H$. The computation of the first
$R(K)-1$ figures would be carried out all right, but the instructions
for calculating the $R(K)$-th would amount to ``calculate the first
$R(K)$ figures computed by $H$ and write down the $R(K)$-th''. This
$R(K)$-th figure would never be found. {\it I.e.,} $H$ is circular,
contrary both to what we have found in the last paragraph and to the
verdict ``$s$''. Thus both verdicts are impossible and we conclude
that there can be no machine D\footnote{A machine that is assumed
capable of determining whether a given machine is circular or
not. This machine is introduced to construct $H$.}.''}\\ (Footnotes
added by the author)
\end{quote}
In this paragraph, Turing considered the {\em actual behavior} of
intact $H$ on its self-description $K$, and noticed that what this
machine would need to compute is exactly the same situation as the
machine itself is in: {\em ``$H$ is looking at its self-description
$K$.''} Such a self-reference would result in a circular process that
never comes back. Therefore, $H$ cannot make any decision on whether
$K$ is satisfactory or not. This contradiction negatively proves the
possibility of $D$, or a general computational procedure to determine
whether a machine stops writing symbols or not.

\section{Self-replication emerging}

Turing's argument described above gives essentially the same argument
as to what could happen if $M'$ in our notation received its
self-description $p(M')$. In this case $M'$ must compute the value of
$f'(p(M')) = f(p(M'),p(M'))$, and hence it would need to compute the
behavior of the machine described in $p(M')$ on the input $p(M')$,
exactly the same circular situation as that appearing in Turing's
proof. Let us use this example in what follows, as it is much simpler
to understand than Turing's original settings.

What kind of computational task would $M'$ be carrying out in this
circular situation? It tries to compute the behavior of $M'+p(M')$,
which tries to compute the behavior of $M'+p(M')$, which tries to
compute the behavior of $M'+p(M')$, ... Interestingly, this chain of
self-computation takes place in the form identical to that of
self-replication in von Neumann's construction theory shown in Eq.\
(\ref{selfrep}), if ``{\em to compute the behavior}'' is read as
``{\em to construct the structure}''. This similarity may be better
understood by noting that the role of $\cal C$ that attaches $\cal
I(X)$ to $\cal X$, shown in the last line of Eq.\ (\ref{roleofC}),
parallels the role of diagonalization in Eq.\ (\ref{diagonal}); {\em
both attempt to apply a copy of the description to the machine
represented by the description.}

\begin{figure}[tp]
\begin{center}
\includegraphics[width=0.8\columnwidth]{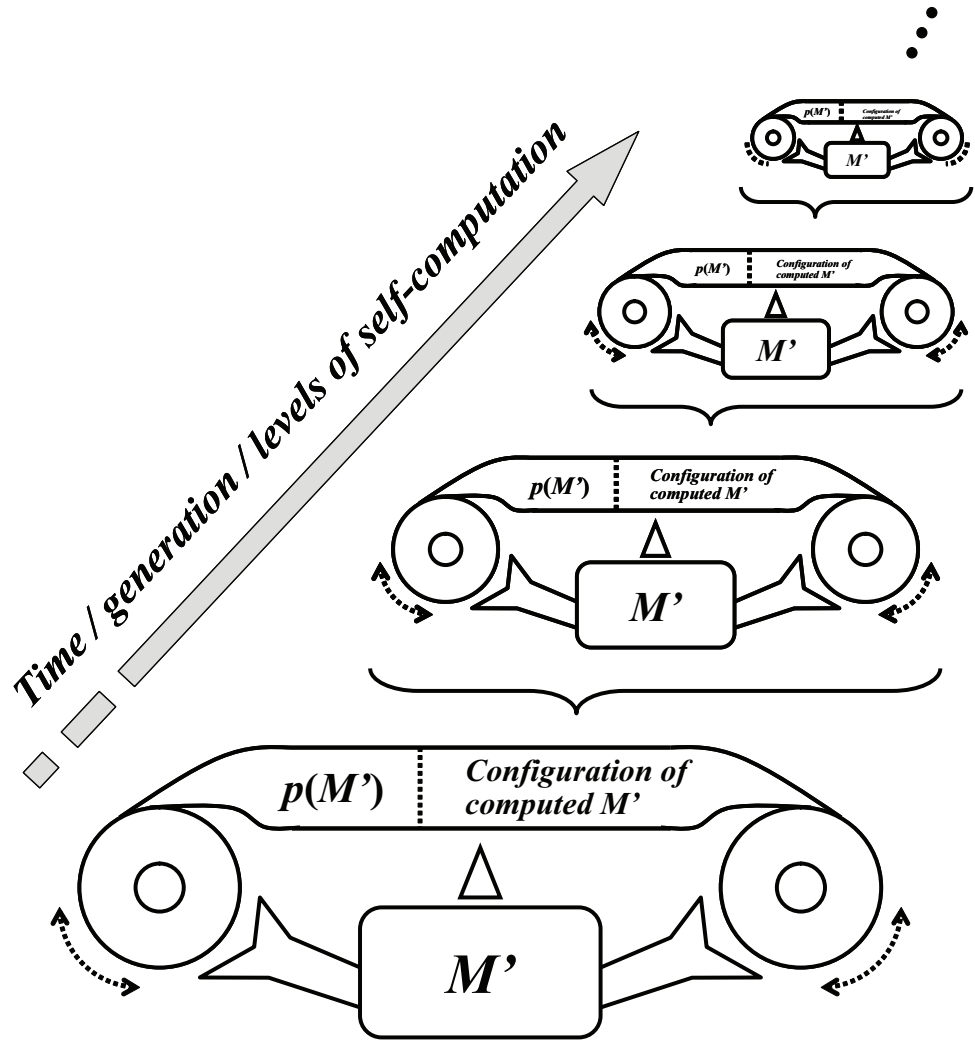}
\end{center}
\caption{Self-replication of Turing machine $M'$ on the tape
space. Given its own description $p(M')$, it starts an indefinite
cascade of self-computation, where the information about $M'$ actually
self-replicates on the tape. The representation of the computed
machine becomes more and more indirect as the level of
self-computation becomes deeper.}
\label{replicating}
\end{figure}

Moreover, if one watched how the actual configuration of the tape of
$M'$ changes during such a self-computing process, he would see that
the information about $M'$ {\em actually self-replicates} on the tape
space, with its representation becoming more and more indirect as the
level of self-computation becomes deeper (Fig.\
\ref{replicating}). Turing might have imagined this kind of
self-replicating dynamics of machines when he developed his argument.

In view of the similarity between the above two processes, it is
clearly recognized that von Neumann's design of self-reproducing
automata is by no means just an anomaly in construction
theory. Rather, it correctly reflects the diagonal situation leading
to an infinite self-computation chain of computationally universal
machines, which appears in the proof of the undecidability of the
halting problem presented by Turing.

\section{Related work}


A well-known argument on the computational undecidability related to
self-replication was developed by Cohen \cite{cohen1987}, where he
showed that there is no general algorithm for the detection of
self-replicating computer viruses. The proof is rather simple: If
there were an algorithm, say $S$, that can determine whether a given
computer program is self-replicative, then one could easily create
another contradictory program that has $S$ built in it and
self-replicates if and only if its $S$ classifies the program itself
as non-self-replicative. This is probably the best acknowledged
discussion on the relationship between self-replication and the
undecidable problem so far.

We should note, however, that Cohen's argument suggests that detecting
a computer program that does {\em ``X''} is generally impossible,
where {\em ``X''} could be self-replication but could also be replaced
by any other functions; self-replication is no more than just one of
many possible behaviors of universal machines. In contrast, our
argument discussed in this essay is more fundamental: Universal
machines may fall into undecidable situations {\em because of the
possibility of self-replication (either computation or construction).}
Here self-replication is not just an instance of many possible
behaviors, but is actually the key property that causes the
undecidability of the behavior of universal machines, either
computational or constructional.

Another related work would be the theory of self-replicative recursive
functions discussed in recursion theory in 1960's, where Kleene's
recursion theorem \cite{kleene1938} was applied to prove that there
exist recursive functions (i.e., computer programs) that produce their
own representations as outputs, regardless of given inputs
\cite{myhill1964,rogers1967}. These functions were later implemented
as actual computer programs and named {\em quines}
\cite{hofstadter1979}; writing quines in various programming languages
has been one of the standard amusements in computer science
community. A common way of creating a quine is to embed a partial
representation of the program in itself and use it twice for creating
a main body of the program and for re-embedding a copy of the
representation into the newly created program. This technique of {\em
quining} is exactly the same as what von Neumann proposed in his
formulation of self-replicating machines.

There is, however, at least one fundamental difference between these
self-replicating programs in recursion theory and the self-replicating
constructors in construction theory. In the former case, the
computation process always stops after producing a static
representation of the program, with no direct implication derived on
its relevance to the halting problem. In the latter case, on the other
hand, the construction process never stops because the product of
construction, as active as its constructor, starts its own
construction process once fully constructed.

Interestingly, Turing's argument in his proof of the undecidability of
the halting problem considered the {\em latter} case in computation
theory, where the Turing machine is not simply writing its own
representation (as usual quine programs do), but is actually trying to
compute its own dynamic behavior (as illustrated in Fig.\
\ref{replicating}). This point may be well understood by reminding
that a product being constructed in construction theory corresponds
not to symbols being written, but to a {\em computational process
itself}, in computation theory.

\section{Conclusion}

As Turing showed in his proof, when a computational machine tries to
solve the halting problem of its own computation process, it will fall
into a cycle of self-computation that never ends in a finite time. Our
point is that this corresponds exactly to the cycle of
self-replication in construction theory, and that von Neumann's
self-reproducing automaton model rightly captures this feature in its
formulation. The halting problem solver in construction theory lets
the subject machine construct its product and see if it eventually
stops. If it tries to solve the halting problem of its own
construction process, it will start self-replication, and the entire
process never completes in a finite amount of time.

The insight obtained in the above sections provides us with some new
implications about the connections between computation and
construction. Throughout our argument, we saw that the construction of
another machine in construction theory has the same role and meaning
as does the computation of another machine in computation theory. This
correspondence transcends the second difference between computational
and constructional machines we discussed in the Introduction, where I
said:
\begin{quote}
{\em The computational universality is defined by the ability of
computing the behavior of all the other models of computation, while
the constructional universality is defined by the ability of
constructing all the structures in a given specific product set, which
has nothing to do with the ability of computing the behavior of other
constructors.}
\end{quote}
Interestingly, once construction and computation are identified with
each other, these two universalities become very close---if not
exactly the same---so that the universal constructor indeed has the
ability to compute the behavior of all the other constructors {\em by
physically constructing them and letting them do their jobs}. The idea
of such ``constructor-constructors'' is relevant to the realization of
machines with epigenetic dynamics, which will be one of the more
important subjects in construction theory.

The computation-construction correspondence also gives us a unique
view of biology, suggesting that the relationship between parent and
offspring in biological systems is equivalent to the relationship
between the {\em computing} $M'$ and the {\em computed} $M'$ in
computation theory. From a construction-theoretic perspective, a
biological organism is trying to find out the final result of the
construction task written in its genotypic information by executing
its contents. The final product will be immediately found if the
product of the task is a static structure, such as drugs produced by
genetically modified bacteria. But if the product is another active
machine that will attempt to build other products, then the final
result will depend on what this product will do next. Furthermore, if
the product is identical (or sufficiently similar) to the original
organism itself, the situation represents the conventional
parent-offspring relationship, where offspring are a kind of
intermediate product produced during the whole long-standing
construction process.

In this view, the endless chain of self-replication that living
systems are in, may be reinterpreted as a parallel to the endless
chain of self-computation that a halting problem solver falls in. In a
sense, we may all be in the process initiated billions of years ago by
a first universal constructor, who just tried to see the final product
of its {\em diagonal} construction.

\section*{Acknowledgments}

I would like to thank William R.\ Buckley for his continuous
encouragement and helpful suggestions, and also four anonymous
reviewers who gave me very constructive and insightful comments that
significantly improved the quality and correctness of the ideas
presented here.

\end{document}